\documentclass[aps,prb,twocolumn,showpacs,superscriptaddress]{revtex4-1}
\usepackage{graphics,graphicx}
\usepackage{color}
\usepackage{amssymb,amsmath}
\usepackage{bm}
\usepackage{mhchem}
\usepackage[normalem]{ulem}  
\usepackage{comment}
\usepackage[caption=false]{subfig}

\usepackage {hyperref}
\hypersetup{
    colorlinks=true,
    linkcolor=blue,
    citecolor=blue,
    linktoc=all
}

\begin{document}
\title{Dimer-induced heavy-fermion superconductivity in the Shastry-Sutherland Kondo lattice model}
\author{Lei~Su}
\affiliation{Centre for Advanced 2D Materials and Graphene Research Centre, National University of Singapore, 6 Science Drive 2, Singapore 117546}
\author{Pinaki~Sengupta}
\affiliation{School of Physical and Mathematical Sciences,\\ Nanyang Technological University, 21 Nanyang Link, Singapore 637371}
\affiliation{Centre for Advanced 2D Materials and Graphene Research Centre, National University of Singapore, 6 Science Drive 2, Singapore 117546}
\date{\today}

\begin{abstract}
We study the  Kondo lattice model on the geometrically frustrated Shastry-Sutherland lattice
focusing on the quantum phase transition between the valence bond solid and the heavy fermion 
liquid phase. By explicitly including spinon pairing of local moments at the mean-field level, 
we establish the emergence of a unique heavy fermion superconducting phase induced by dimer 
ordering of the local moments coexisting with Kondo hybridization. Furthermore, we demonstrate that for suitable choices of parameters, a partial
Kondo-screening phase, where some of the valence bonds are broken, precedes the aforementioned dimer-induced superconducting phase. Our 
results have important implications in understanding the experimental observations in the 
heavy fermion Shasstry-Sutherland compounds. 
\end{abstract}
\maketitle
\newpage
\section{INTRODUCTION}

Heavy fermion materials have been studied extensively ever since they were discovered in the 1970s (For reviews, see Refs.~\onlinecite{ stewart1984heavy,coleman2007heavy}). The underlying physics has long been thought to be captured by the Kondo lattice model in which the Ruderman-Kittel-Kasuya-Yosida (RKKY) interaction between local magnetic moments competes with the Kondo interaction between itinerant electrons and local moments. The competition produces the well-known Doniach diagram.\cite{doniach1977kondo} The diagram is characterized by a continuous phase transition from a long-range magnetically ordered phase with a small Fermi surface to a heavy fermi liquid (HFL) with a large Fermi surface due to liberation of local moments. The transition point is believed to be a prototypical quantum critical point (QCP) that involves large collective quantum fluctuations. In recent years, quantum criticality in heavy fermion metals has evolved to a central subject to study unconventional phases in strongly correlated electron systems.\cite{coleman2005quantum,gegenwart2008quantum}

However, experimental observations of separation of the QCP from the small-large Fermi surface reconstruction, e.g. for YbRh$_2$Si$_2$ under pressure or doping, \cite{tokiwa2009separation, friedemann2009detaching,custers2010evidence} strongly suggest that the Doniach conceptual picture may be incomplete. Frustration, as a source of many exotic phases, \cite{balents2010spin} is believed to have an important role in shaping the global phase diagram of various Kondo lattices. \cite{si2010quantum,coleman2010frustration} In fact, different groups of frustrated heavy fermion materials have been discovered in experiments. For instance, pyrochlore Pr$_2$Ir$_2$O$_7$ is geometrically frustrated\cite{nakatsuji2006metallic} while tetragonal CeRhIn$_5$ is frustrated magnetically because of interactions between next-nearest neighbors.\cite{park2008isotropic}

\begin{figure}[t]
\centering
\vspace*{-20pt}
\subfloat{
                \includegraphics[width=0.238\textwidth]{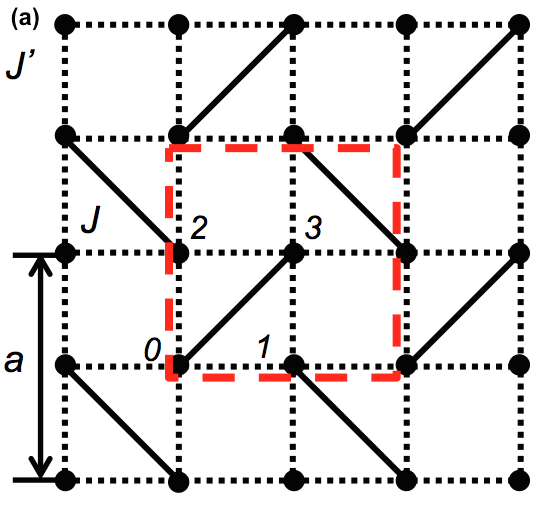}
                \label{fig:SSL}  } 
\subfloat{
        
                \includegraphics[width=0.228\textwidth]{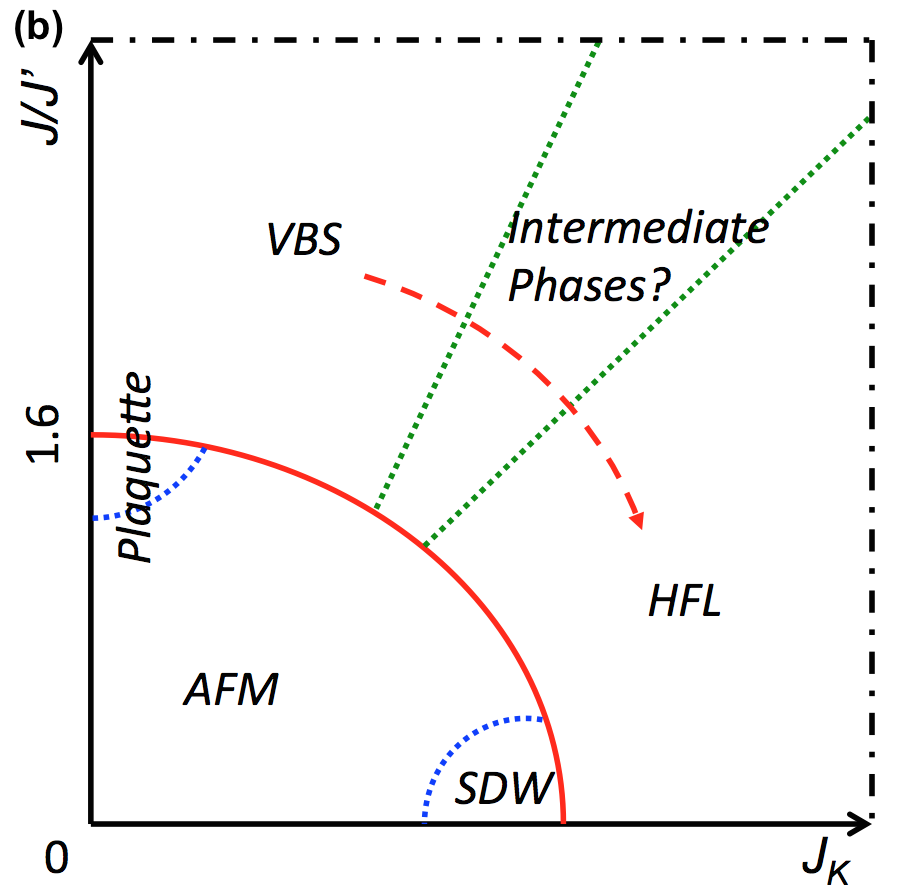}     
                \label{fig:SPD} }         

	\caption{(Color online) (a) Topology of the Shastry-Sutherland lattice. A unit cell, of size $a$, contains four sites and two types of bonds: diagonal bonds $J$ and axial bonds $J'$; (b) Proposed global phase diagram for a Shastry-Sutherland Kondo lattice. The VBS-HFL transition will be our focus.
}   
\label{fig:Schematic}   
\end{figure}

A pair of recently discovered frustrated systems, Yb$_2$Pt$_2$Pb and its cerium analogue, also demonstrate heavy-fermion properties. \cite{kim2008yb, kim2013spin, kim2011heavy} In these materials, 
magnetic moments, with spin 1/2, arise from the $f$ electrons of Yb or Ce atoms localized 
on a Shastry-Sutherland lattice (SSL). \cite{sriram1981exact}  An SSL 
(Fig.~\hyperref[fig:Schematic]{\ref{fig:Schematic}(a)}) is topologically equivalent to 
a square lattice with alternating diagonals. The name derives from the study of the spin-1/2
antiferromagnetic (AFM) Heisenberg model on this lattice by Shastry and Sutherland. They 
showed that the ground state of the Shastry-Sutherland model (SSM) undergoes an interesting 
phase transition as the ratio of the Heisenberg interaction along the diagonal bonds ($J$) 
and that along the principal axes ($J'$) is varied. For small $J/J'$, the ground state is 
gapless with long range AFM order, whereas for $J/J' \gtrsim 1.6$, a gapped valence bond 
solid (VBS) -- consisting of singlets along every diagonal -- is stabilized.\cite{sriram1981exact} 
The model has attracted extensive interest in the recent past following the discovery of 
field-induced magnetization plateaus (akin to quantum Hall plateaus) in \ce{SrCu_2(BO_3)_2}.
\cite{Miyahara1999,Kodama2002,Kageyama1999,Miyahara2003,gaulin2004,Takigawa2008,Misguich2001,
Sebastian2008}. Yb$_2$Pt$_2$Pb, in which the heavy fermion feature suggests the Kondo coupling with the interacting local moments, serves as a 
representative example of frustrated heavy fermion metals and an excellent platform to study the 
effect of frustration on the Doniach phase diagram.

The Shastry-Sutherland Kondo lattice (SSKL) was discussed qualitatively in 
Ref.~\onlinecite{coleman2010frustration} and a schematic ground state phase diagram 
similar to Fig.~\hyperref[fig:Schematic]{\ref{fig:Schematic}(b)} was shown. When 
both the Kondo coupling $J_K$ and the ratio $J/J'$ are small, the system should 
be an antiferromagnet (AFM) as expected for an antiferromagnetic Heisenberg model. 
Possible intermediate phases, such as a plaquette phase between the AFM and the 
VBS phase, and a spin density wave (SDW) phase between the AFM and the HFL phase, 
have been actively studied previously. The small-large Fermi surface transition 
from the VBS phase to the HFL phase, however, has remained largely unexplored. In 
Ref.~\onlinecite{coleman2010frustration}, the authors argued that, in the metallic 
case, both small and large Fermi surface phases may be unstable to $d_{xy}$-wave superconductivity and that it is possible  the two superconducting phases are 
connected smoothly. Unlike the magnetically mediated heavy fermion superconductivity at the antiferromagnetic QCP, \cite{gegenwart2008quantum}  which has been actively studied, these 
proposed superconducting states are induced by the dimer order on the SSL. This 
proposal will be the main focus of our work. The system was also studied in 
Ref.~\onlinecite{bernhard2011frustration} and \onlinecite{pixley2014quantum} 
where the focus was on the global phase diagram and the Heisenberg interaction was 
decoupled only in the valence bond channel. In particular, 
Ref.~\onlinecite{pixley2014quantum} hints at the existence of a partial Kondo 
screening (PKS) phase in a large-$N$ limit. We would like to see how the 
superconducting phase survives the competition with this possible phase.

In the following section, we will first introduce the SSKL model Hamiltonian 
and the mean-field theory that explicitly includes also the decoupling in the 
spinon pairing channel. In Sec.~\ref{sec:Results}, we present our quantitative 
results and show explicitly the existence of both a partial Kondo-screening phase and a dimer-induced heavy fermion liquid phase . In the last section, we discuss the 
implications of our results and summarize our work.

\section{MEAN-FIELD THEORY}
There are three components in the SSKL model and the general Hamiltonian\cite{coleman2010frustration} can be written as
\begin{equation}
H = H_t + H_K + H_{SS}.
\label{SSK} 
\end{equation}
The first term describes the tight-binding hopping of conduction electrons 
on the SSL,
\begin{equation}
H_t = - \sum_{( i,j), \sigma} t_{ij} (c_{i \sigma}^{\dagger}c_{j \sigma} + {\text{H.c.}}).
\label{Ht}
\end{equation} 
We shall assume that the only non-zero hopping matrix elements ($t_{ij}$)
are $t$ for the diagonal bonds and $t'$ for nearest neighbors along
the principal axes (Fig.~\hyperref[fig:Schematic]{\ref{fig:Schematic}(a)}).
The second and the third term describe the Kondo coupling between conduction 
electrons and local moments on the SSL and the Heisenberg interaction between 
local moments, respectively:
\begin{equation}
H_K =  J_K \sum_i {\textbf{s}}_i \cdot {\textbf{S}}_i  
\label{Kondo}
\end{equation}
and
\begin{equation}
H_{SS} =  \sum_{( i,j)} J_{ij}{\textbf{S}}_i \cdot {\textbf{S}}_j,
\label{eq:SSL1}
\end{equation} 
where $J_{ij} = J$ or $J'$ (Fig.~\hyperref[fig:Schematic]{\ref{fig:Schematic}(a)}). 
All interactions are chosen to be anti-ferromagnetic, i.e., $J_K,J,J' \geq 0$. In 
Eq. (\ref{Kondo}), we have described the spin density of conduction electrons in 
its spin representation
\begin{equation}
{\textbf{s}}_i= \dfrac{1}{2} (c_{i \alpha}^{\dagger}{\bm \sigma}_{\alpha\beta} c_{i \beta}),
\end{equation}
where ${\bm \sigma}_{\alpha\beta}$ are Pauli matrices and summation over repeated dummy indices is assumed. 

The presence of spin operators in the Hamiltonian makes perturbation theory hard to handle, and in order to obtain a mean-field theory, we use the pseudo-fermion representation of local moments,
\begin{equation}
{\textbf{S}}_i = \dfrac{1}{2} (f_{i\alpha}^{\dagger}{\bm \sigma}_{\alpha\beta} f_{i \beta}), 
\label{eq:Spseudo}
\end{equation}
under the constraint 
\begin{equation}
\sum_{\sigma} f_{i\sigma}^{\dagger} f_{i \sigma} = 1 
\label{eq:fconstraint}
\end{equation}
on each site $i$. This representation has a well-known $SU(2)$ gauge invariance. \cite{affleck19882,coleman1989kondo} 

Making use of the identity
\begin{align}
{\textbf{S}}_i \cdot {\textbf{S}}_j  - \frac{1}{4} &= - \frac{1}{2} ( \sum_{\sigma} f_{i \sigma}^{\dagger} f_{j \sigma})( \sum_{\sigma'}f_{j\sigma'}^{\dagger}f_{i \sigma'}) \nonumber \\ 
&=-\frac{1}{2} ({\cal {\epsilon}}^{\alpha\beta} f_{i\alpha}^{\dagger}f_{j\beta}^{\dagger})({\cal {\epsilon}}^{\gamma\delta} f_{j\delta}f_{i\gamma}) 
\label{eq:Id}
\end{align} 
under the constraint Eq.~(\ref{eq:fconstraint}), we can decouple the four-point operator in two different channels: $\gamma_{ij} =\sum_{\sigma}\langle f_{i \sigma}^{\dagger} f_{j \sigma} \rangle$ and $ b_{ij} =\langle {\cal {\epsilon}}^{\alpha\beta} f_{i\alpha}^{\dagger}f_{j\beta}^{\dagger} \rangle$, where ${\cal {\epsilon}}^{\alpha\beta} $ is the anti-symmetric tensor with ${\cal {\epsilon}}^{\uparrow\downarrow} =1$. In order to study the spinon pairing condensation $b_{ij} \neq 0$  in a gauge invariant manner, we keep the decoupling in both channels explicitly. \cite{coleman1989kondo} The idea can also be justified using the symplectic large-$N$ \cite{flint2008heavy} mean-field theory or the Hartree-Fock approximation \cite{bernhard2011frustration}(with a scaling factor in couplings). A similar decoupling scheme may be carried out for the Kondo interaction in Eq.~(\ref{Kondo}), but we can always remove the pairing $\langle {\cal {\epsilon}}^{\alpha\beta} f_{i\alpha}^{\dagger}c_{i\beta}^{\dagger} \rangle$ through an $SU(2)$ gauge transformation. \cite{flint2008heavy}  Under this gauge-fixing condition, only the Kondo hybridization  $V_i =\sum_{\sigma}\langle f_{i \sigma}^{\dagger} c_{i\sigma} \rangle$ will be considered. Notice that in this case, only the constraint in Eq. (\ref{eq:fconstraint}), instead of the constraint triplet,\cite{coleman1989kondo} needs to be satisfied in order to maintain the gauge invariance of the Kondo coupling. 

Implementing multipliers $E^f_i$ for the constraint of single occupation and $\mu$ for the band-filling
\begin{equation}
 \sum_{\sigma}\langle c^{\dagger}_{i\sigma} c_{i\sigma} \rangle = n_c,
 \label{nc}
\end{equation} 
we are able to arrive at the following Hamiltonian
\begin{align}
\tilde{H} =& E_0 - \sum_{( i,j), \sigma} t_{ij}(c_{i \sigma}^{\dagger}c_{j \sigma} + {\text{H.c.}})  - \mu \sum_{\sigma} c_{i \sigma}^{\dagger}c_{i \sigma} \nonumber \\
&- \frac{1}{2}\sum_{( i,j)} J_{ij} (\sum_{\sigma} \gamma_{ij} f_{j \sigma}^{\dagger}f_{i \sigma}  +b_{ij}  {\cal {\epsilon}}^{\beta\alpha} f_{j\alpha}f_{i\beta}  + {\text{H.c.}} ) \nonumber \\ 
& +\sum_{\sigma}E^f_i f_{i \sigma}^{\dagger}f_{i \sigma} -  \frac{1}{2}\sum_{i,\sigma} J_K (V_i c_{i \sigma}^{\dagger} f_{i \sigma} + {\text{H.c.}} ) ,
\label{Eq: Hamiltonian2}
\end{align}
where 
\begin{equation}
E_0 =  \sum_{( i,j)} \frac{J_{ij}}{2} (|\gamma_{ij}|^2 +|b_{ij}|^2  )  +  \sum_i \frac{J_K}{2} |V_i|^2 + \sum_i (\mu n_c -E^f_i).
\end{equation}
We can break the $Z_2$ symmetry of spins in order to incorporate the AFM phase as in Ref.~\onlinecite{bernhard2011frustration}, but we exclude them here for simplicity. 

Assuming translational symmetry allows us to take $V_i =V$, $E^f_i = E_f$, and $\gamma_{ij} = \gamma, \gamma', |b_{ij}| = b, b'$, respectively, depending on the type of the bond $(i,j)$.  $V_i, \gamma_{ij}$ can be made real through gauge transformations, but $b_{ij}$ is not real in general. The single occupation constraints, however, force spinon pairs to condense $d_{xy}$-symmetrically:\cite{coleman2010frustration} $b_{i+x,i} = - b_{i, i+y} $ and $b_{i+x,i+y} = - b_{i-x, i+y} $, where $ x, y = a/2$. We will see soon that the PKS phase, as a phase other than the VBS that breaks the translational symmetry, may also be a ground state, and in that case, order parameters will be more inhomogeneous.

Written in terms of Nambu spinors, the mean-field Hamiltonian can be expressed as
\begin{equation}
{\tilde{H}}_{MF} = \sum_{ \textbf{k}} {\Psi}^{\dagger}(\textbf{k}){\tilde{T}}(\textbf{k}) {\Psi} (\textbf{k})  + \tilde{E}_0 ,
\end{equation}
where ${\tilde{T}}(\textbf{k})$ are $16 \times 16$ symmplectic matrices (See Appendix) and 
\begin{equation}
\tilde{E}_0 =  \sum_{( i,j)} \frac{J_{ij}}{2} (|\gamma_{ij}|^2 +|b_{ij}|^2  )  +  \sum_i \frac{J_K}{2} |V_i|^2 + \sum_i \mu (n_c -1).
\label{Eq:tE0}
\end{equation}
The eigenvalues of ${\tilde{T}}(\textbf{k})$ come in positive-negative pairs, and the ground state energy per site can, therefore, be written as
\begin{equation}
E  =  ( \sum_{\textbf{k} \nu} E_{\nu}^{-}(\textbf{k}) + \tilde{E}_0) /4N,
\end{equation}
with $\nu$ denoting band index, $N$ the number of unit cells, and $\textbf{k}$ runs over the first Brillouin zone. Minimizing the average ground state energy under the constraints Eqs. (\ref{eq:fconstraint})(\ref{nc}) allows us to find the self-consistent mean-field order parameters and to determine the ground state phases.

Notice that once the spinon pairs condense, the simultaneous condensation of the Kondo hybridization implies an induced instability in conduction electron pairs.\cite{coleman1989kondo,senthil2003fractionalized, liu2012d} In our mean-field theory, it can be approximated by
\begin{equation}
\Delta_{ij} = |\langle \epsilon^{\alpha\beta} c^{\dagger}_{i\alpha}c^{\dagger}_{j\beta}\rangle |=   \frac{ 2 K^2 B_{ij}}{U\sqrt{(U +W)}},
\label{eq:Delta}
\end{equation}
where
\begin{equation}
U = 2\sqrt{|B_{ij}|^2 \mu^2 + (|K|^2 + E_f \mu)^2},
\end{equation}
and
\begin{equation}
W = (E_f^2 +\mu^2 + |B_{ij}|^2) +2 |K|^2.
\end{equation}
with $B_{ij}  = J_{ij} b_{ij}/2$ and $K = J_K V/2$. By replacing the order parameters on diagonal bonds with the corresponding ones on axial bonds, we also have similar expressions for the condensation on the principal axes. The Eq.~(\ref{eq:Delta}) has inspired the authors of Ref.~\onlinecite{coleman2010frustration} to postulate that such an induced superconducting phase may dominate the zero temperature phase diagram of the SSKL model. In the following section, we will show that such a phase is indeed possible but is restricted by multiple factors.

\begin{figure}[b]
  \includegraphics[width=0.45\textwidth]{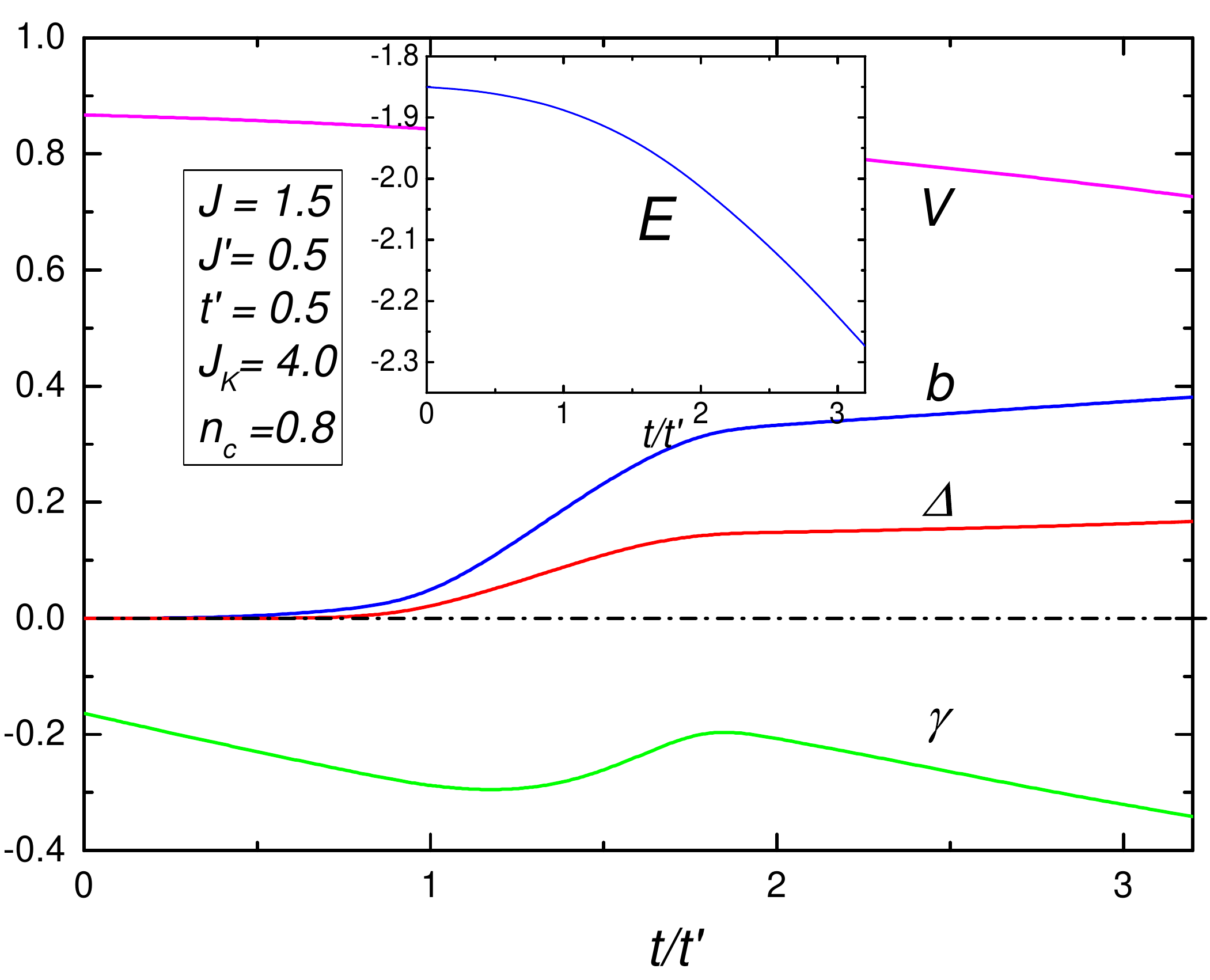}   
       \caption{(Color online) Mean-field order parameters in the $d_{xy}$-HFSC phase as a function of hopping ratio, $t/t'$, when $t'$ is large and $n_c$ close to half-filling. The valence bond order along the diagonal bonds, $\gamma$, is always non-zero and the corresponding spinon pairing amplitude $b$ rises quickly only after the diagonal hopping $t$ becomes comparable with the axial hopping $t'$. The induced pairing of conduction electrons $\Delta$ is also shown. The inset shows the ground state energy.
}   
\label{Graph1}
\end{figure}

\section{RESULTS}
\label{sec:Results}

First, we examine the proposed coexisting state, focusing on possible intermediate phases between the VBS and the HFL phase. Apart from the VBS phase, let us first consider only phases that do not break the translational symmetry of the lattice. We emphasize that both valence bond orders $\gamma_{ij}$ and spinon pairing condensations $b_{ij}$ are considered simultaneously since the condensation in the Kondo channel will spontaneously break the gauge invariance. At half-filling, there is only one direct first order transition from the VBS to the HFL phase. In fact, the latter is the so called Kondo insulator phase. When the band-filling is lowered, we find that the ground state is still a VBS phase at a small $J_K$. When $J_K$ is increased, there is a first order phase transition from the VBS state to either a phase with dimer-induced conduction electron pairing condensation which we will call a heavy fermion superconducting ($d_{xy}$-HFSC) phase henceforth, or a normal heavy Fermi liquid (HFL). If we lower $J$ simultaneously, in both cases, we can see a transition from the $d_{xy}$-HFSC or the HFL phase to a Fermi liquid with a $d_{x^2 -y^2}$-wave superconducting instability, which phase was discussed previously.~\cite{coleman1989kondo,liu2012d} In other words, the intermediate phase in Fig.~\hyperref[fig:Schematic]{\ref{fig:Schematic}(b)} can be either a $d_{xy}$-HFSC phase or a normal HFL phase. 

\begin{figure}
  \includegraphics[width=0.45\textwidth]{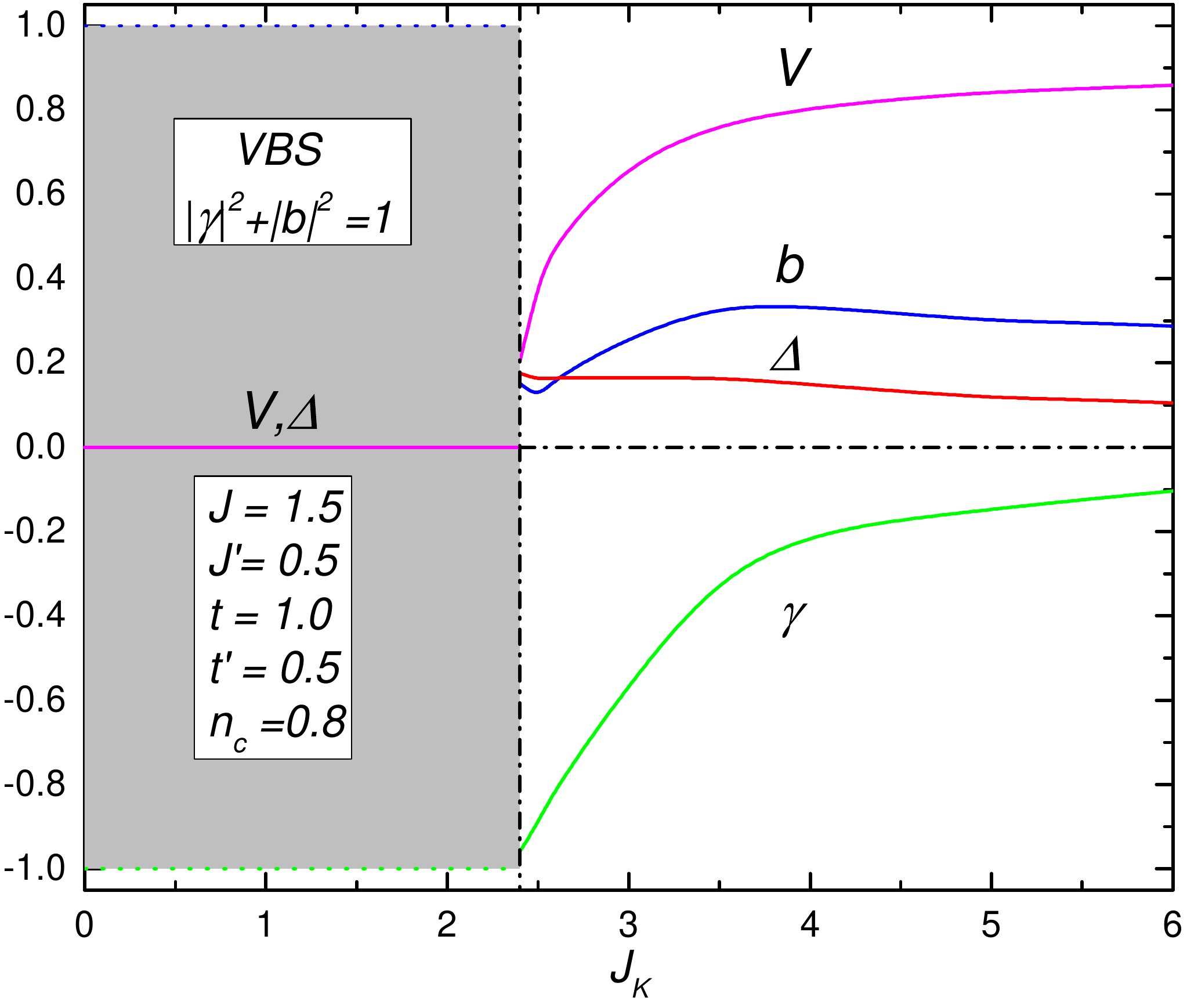}   
  \vspace*{-5pt}
       \caption{(Color online) Mean-field parameters as a function of the Kondo coupling $J_K$. In the VBS phase, the order parameters $\gamma$ and $b$ satisfy the relation $|\gamma|^2 +|b|^2 =1$ due to the gauge invariance. The transition at around $J_K =2.4$ is of first order.
}   
\label{Graph3}
\vspace*{-10pt}
\end{figure}

In either of these two phases, the order 
parameters $\gamma_{ij}$ and $b_{ij}$ subtly balance to minimize the mean-field ground 
state energy. Apart from coupling strengths and the band-filling of conduction electrons $n_c$, the balance may also depend on the hopping amplitudes, $t$ and $t'$, along the diagonal and axial bonds, respectively. This is true when the band energy due to hopping is comparable to the other two components at electron densities close to half-filling. A typical dependence in this case is shown in Fig.~\ref{Graph1} at $n_c =0.8$. Notice that only the order parameters 
on the diagonal bonds are shown as spinon pairing $b'$ on the axial bonds always vanishes 
and $\gamma'$ is relatively insignificant. When $t$ is smaller than $t'$, the spinon 
pairing amplitude $b$, hence the condensation amplitude $\Delta$ of conduction electrons, 
is small. Only when $t$ becomes comparable with $t'$ does the amplitude rise to the same 
order as $\gamma$. However, we do not observe a strict separation between the two phases. Meanwhile, we find that when either $t'$ and $t$ are very small or the conduction electrons are far from half-filling, e.g., $n_c =0.5$,  the exact value of $t'$ or $t$ is not important. Due to these observations, we conclude that it is $J/J'$ and $J_K$ that play a major role in shaping the phase diagram when no other intermediate phase is considered. In the following analysis, we will focus on the case when the intermediate phase is superconducting.

\begin{figure}
        \centering
        \subfloat{
                \includegraphics[width=0.45\textwidth]{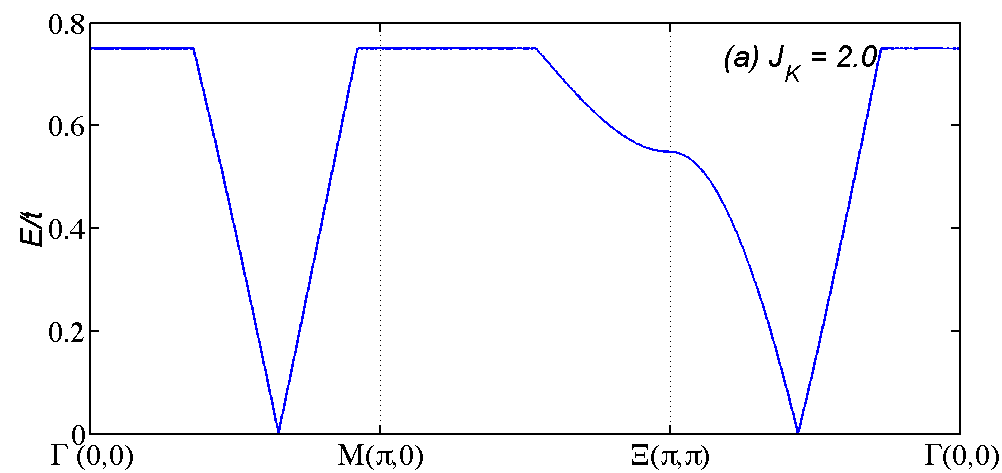}
                \label{fig:K2}       }\\
        
        \subfloat{
                \includegraphics[width=0.45\textwidth]{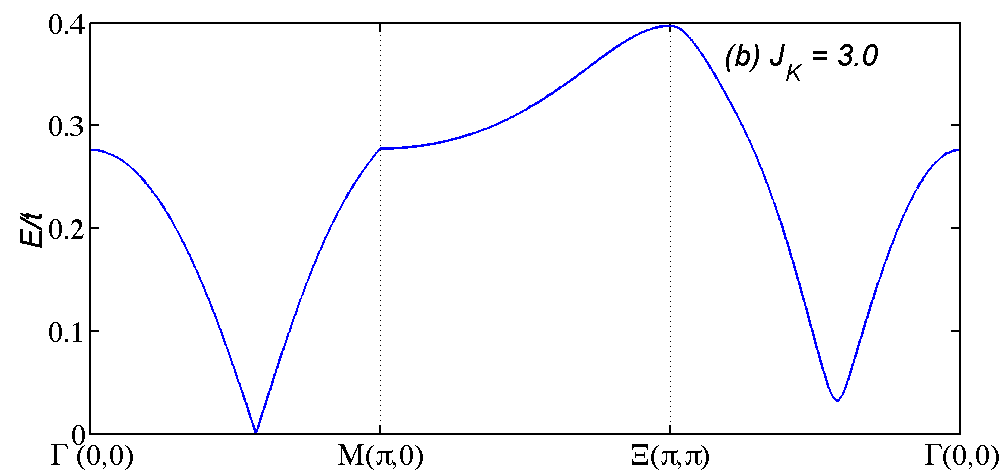}
                \label{fig:K3}       }\\
       \subfloat{
                \includegraphics[width=0.45\textwidth]{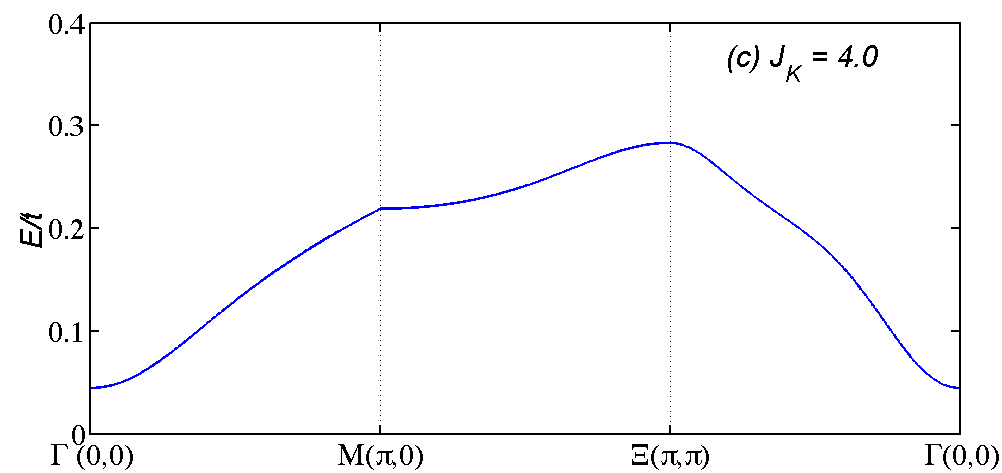}
                \label{fig:K4}       }               
       
       \caption{(Color online) Lowest excitation dispersions for different Kondo couplings $J_K$ =2.0, 3.0, and 4.0, with $J =1.5, J' =0.5, t =1.0, t' =0.5$, and $n_c =0.8$.
}   
\label{GraphK}
\vspace*{-10pt}
\end{figure}

As an example, a cross-section at $J/J' =3$ of the phase diagram is given in 
Fig.~\ref{Graph3}. We can see that when $J_K$ is small the system is in a VBS 
phase, signaled by the condition $|\gamma|^2 + |b|^2 = 1$ as a direct result of gauge invariance. The 
transition takes place around $J_K =2.4$. The discontinuity in the amplitude of order 
parameters depends on the diagonal hopping $t$ -- larger $t$ leads to smaller 
discontinuities. This is because a larger diagonal hopping lowers the Kondo 
hybridization amplitude. A typical evolution of the dispersion of the lowest
excitation mode is shown in Fig.~\ref{GraphK}. When the system is in a VBS phase 
(Fig.~\hyperref[GraphK]{\ref{GraphK}(a)}), there is no excitation gap as a result 
of the metallic character of the conduction electrons. When the HFSC phase becomes 
the ground state, spinon pairs condense and excitations along directions other than 
$\textbf{k}_x =0$ or $ \textbf{k}_y =0$ become gapped, verifying the $d_{xy}$-wave symmetry 
$\Delta(\textbf{k}) \propto \text{sin}(\frac{\textbf{k}_xa}{2})\text{sin}(\frac{\textbf{k}_ya}{2})\Delta$ 
(Fig.~\hyperref[GraphK]{\ref{GraphK}(b)}). As the Kondo coupling $J_K$ increases 
further, the nodes move to the center of the Brillouin zone, $\Gamma (0,0)$, and the 
excitation then becomes fully gapped (Fig.~\hyperref[GraphK]{\ref{GraphK}(c)}). 
The band widths clearly show that the quasi-particle excitations become 
heavier as the Kondo coupling turns stronger.

\begin{figure}
  \includegraphics[width=0.4\textwidth]{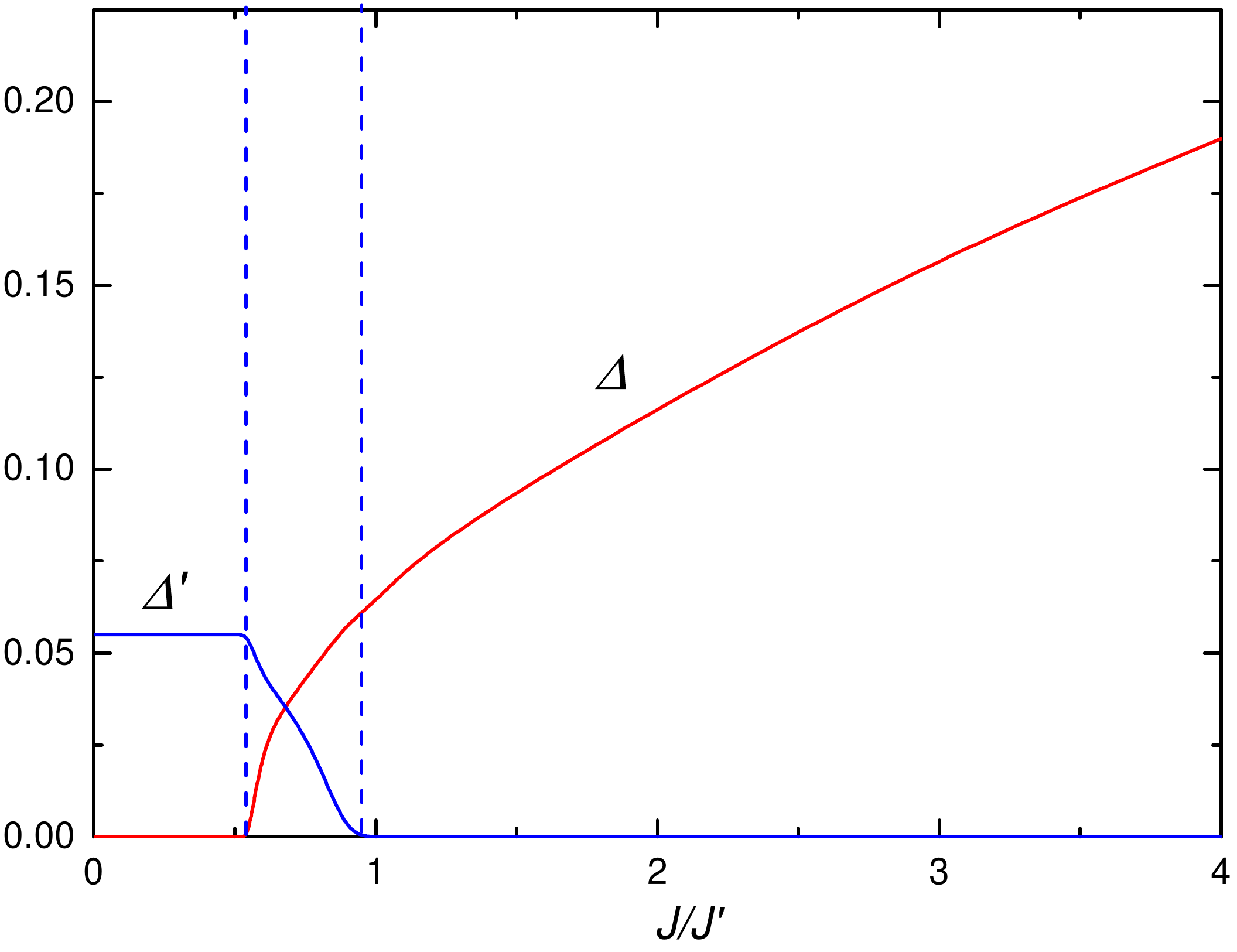}   
       \caption{(Color online)  Condensation of conduction electrons, $\Delta$ and $\Delta'$, as a function of $J/J'$ with $t' =0.5, t =3t', J'=0.5, J_K =4t$ and $n_c =0.5$. The amplitude, $\Delta$, goes to zero at around $J/J' \approx 0.53$ and $\Delta'$ develops from $J/J' \approx 0.95$. Between these two values, the orders coexist. Both transitions are continuous.
}   
\label{Graph9}
\vspace*{-10pt}
\end{figure}

Let us consider the situation where we decrease the ratio $J/J'$ when the system is in the $d_{xy}$-HFSC phase. In the limiting case, $J =0$, the model reduces to a Kondo lattice model on a square lattice, and we expect the ground state phase to be either a HFL or a superconducting phase with $d_{x^2-y^2}$ symmetry when the conduction band is not at half-filling. In fact, within the mean-field formalism, we found that spinon pairing condensation can lower the total energy at small $J/J'$, as is consistent with the previous work.~\cite{coleman1989kondo,liu2012d} A typical example is shown in Fig.~\ref{Graph9}. We see that the diagonal order $\Delta$ goes to zero smoothly as we lower $J/J'$ while the axial condensation $\Delta'$ rises from zero continuously. Once $\Delta$ vanishes, the magnitude of $\Delta'$ becomes almost stable. There is range where both orders coexist.

So far, we have only considered a direct transition from the VBS phase to a $d_{xy}$-HFSC 
or a HFL phase. Can there be any other possible intermediate phase? The answer is 
affirmative. A PKS phase in geometrically frustrated Kondo lattices was first 
discussed in Ref.~\onlinecite{motome2010partial}. In this phase a finite fraction 
of the local moments are still ordered while the rest form singlets with conduction 
electrons. The PKS in the SSKL model was recently discussed in 
Ref.~\onlinecite{pixley2014quantum} in a metallic phase using a large-$N$ mean-field theory. As a result of the restriction of our mean-field formalism, the phase for $n_c =0.5$ is the easiest to discuss as then half of 
the valence bonds on the diagonals will still be locked while the other half are broken 
as the Kondo coupling increases. Our mean-field theory is consistent with the result 
in Ref.~\onlinecite{pixley2014quantum} because at $n_c =0.5$, one sublattice is exactly 
filled, and the Kondo hybridization in the PKS phase are stabilized solely by the 
valence bond order $\gamma$, similar to a Kondo insulator at half-filling.

\begin{figure}[b]
  \includegraphics[width=0.45\textwidth]{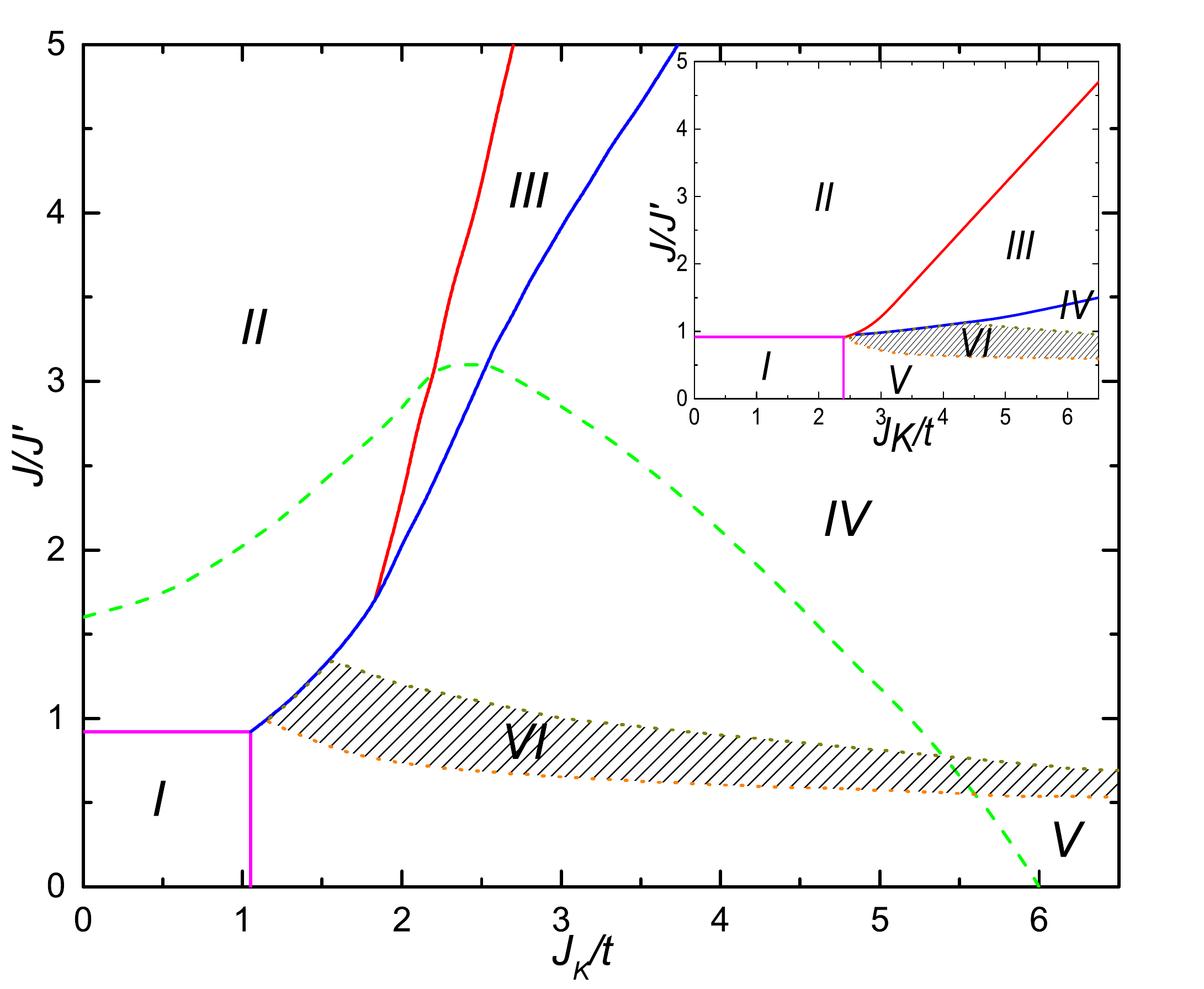}   
       \caption{(Color online) Global phase diagram for $t' =0.5, t =3t', J'=0.5$ and $n_c =0.5$. All the solid boundaries represent first order transitions at the mean-field level approximation. Phase notations: I, P-VBS; II, D-VBS; III, PSK; IV, $d_{xy}$-HFSC; V, $d_{x^2-y^2}$-HFSC; VI (shaded area), coexisting of IV and V. Dotted curves represent continuous transitions. The dashed line denotes the boundary obtained using the approach in Ref. \onlinecite{bernhard2011frustration}. The inset shows the quantitative shift of the boundaries when the hoppings are lowered, $t' =0.1, t = 3t'$, while keeping all others unchanged. 
}   
\label{Graph2}
\vspace*{-10pt}
\end{figure} 
To complete the phase diagram, we need to break the $Z_2$ symmetry of 
the spins to produce the AFM phase -- the ground state is known to possess 
long-range AFM order at small $J/J'$ and $J_K/t$. However, $Z_2$-symmetry 
breaking AFM order parameter is not captured with the framework of the 
symplectic large-$N$ mean field theory that we have used in this study. Without breaking the $Z_2$ symmetry 
explicitly, we obtain a plaquette valence bond solid (P-VBS) phase where only 
$b$ and $V$ vanishes but not the axial $b'$, consistent with the previous 
result.\cite{pixley2014quantum} Since the AFM phase is completely suppressed at
large $J/J'$ and/or large $J_K/t$, this shortcoming does not alter the
central results of our study.  Alternative mean field theories have been used
to capture the AFM phases \cite{bernhard2011frustration,pixley2014quantum} but
they suffer from the inability to describe reliably the PKS and HFSC phases. The 
ground state phase diagram  is shown in Fig.~\ref{Graph2}, for a representative
set of parameters with $t/t' =3$ and $n_c =0.5$. The effect of varying the parameters
on the ground state phases is shown in the inset.  The dashed line indicates the AFM 
phase boundary obtained using the mean-field method given in 
Ref.~\onlinecite{bernhard2011frustration}. It should be noted that this approach 
overestimates the extent of the AFM phase. To distinguish from the P-VBS (I), we denote the dimer phase for a SSL as D-VBS (II) in Fig.~\ref{Graph2}. 
We have also used $d_{xy}$-HFSC (IV) and $d_{x^2-y^2}$-HFSC (V) to distinguish the 
two possible heavy fermion superconducting phase. The range where both 
condensation order parameters are nonzero is indicated by the shaded area (VI). 
As $J$ goes to zero, the SSL characteristic gradually disappears 
and the $d_{xy}$-HFSC phase evolves to a $d_{x^2-y^2}$-HFSC phase.  

Note that the hopping energy has a strong influence on the boundary between the 
PKS (III) phase and the $d_{xy}$-HFSC phase. On the one hand, when the hopping energy is large, the kinetic energy cost
due to the formation of Kondo singlets with local moments is significant. On the other 
hand, when the $d_{xy}$-HFSC phase takes over the PSK phase, the Kondo hybridization amplitude 
will be reduced because the number of local moments coupling to the conduction electrons 
is doubled. Only when the formation of Kondo singlets is able to compensate the cost in 
the valence bond resolution and the cost in kinetic energy of conduction electrons will 
the $d_{xy}$-HFSC phase become the ground phase. As a result, for fixed $J/J'$, the ground state phase 
will transit from the PKS phase to the $d_{xy}$-HFSC phase at a relatively small $J_K/t'$ only when the hopping energy is comparable to the other two components. This effect can be seen explicitly from the inset of Fig.~\ref{Graph2}, where the hoppings are reduced to 1/5 of the initial values.

\begin{figure}[hd]
 \centering
        \subfloat{
  \includegraphics[width=0.45\textwidth]{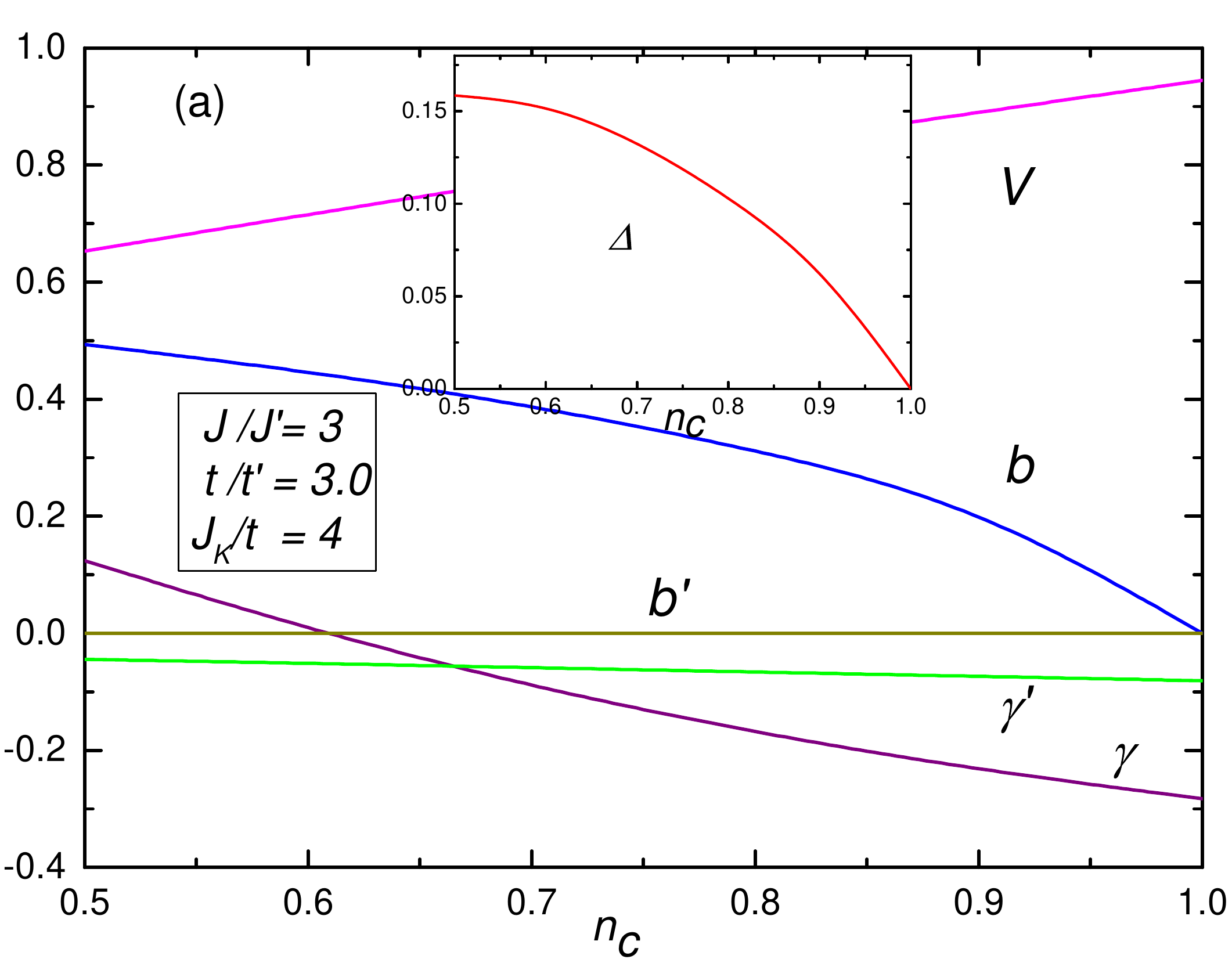}     }\\
  \vspace*{-8pt}
     \subfloat{
\includegraphics[width=0.45\textwidth]{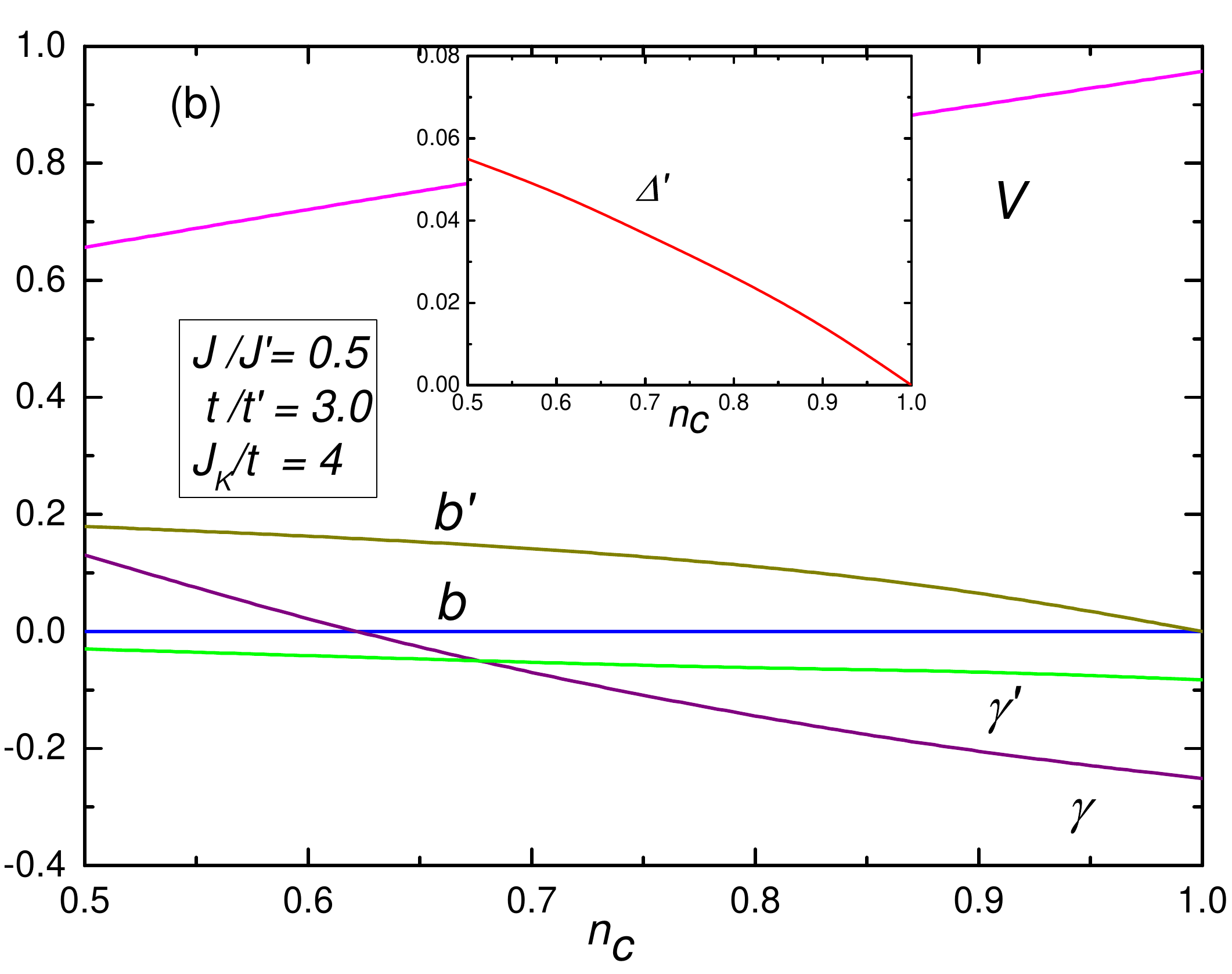}
\vspace*{-8pt}   
}  
 \caption{(Color online) Mean-field order parameters as a function of band filling $n_c$ in a $d_{xy}$-HFSC phase (a) and a $d_{x^2-y^2}$-HFSC phase (b). The insets show the induced pairing of conduction electrons. When $n_c =1.0$, the system becomes a so-called Kondo insulator.
}   
\label{Graphnc}
\vspace*{-10pt}
\end{figure}

In Fig.~\hyperref[Graphnc]{\ref{Graphnc}(a)}, we show the $n_c$-dependence of order 
parameters at $J/J' =3$ and $J_K/t =4$ for which the ground phase 
is $d_{xy}$-HFSC at$n_c =0.5$, as can be seen from Fig.~\ref{Graph2}. At half-filling, the 
system is a Kondo insulator, and its spinon pairing amplitude $b$ of local moments 
vanishes while the finite (negative) valence bond order $\gamma$ stabilizes 
the Kondo hybridization. \cite{iglesias1997revisited}  For comparison, we also show the evolution of order parameters for small $J/J' = 0.5$  in Fig.~\hyperref[Graphnc]{\ref{Graphnc}(b)}. For this set of parameters, the 
system approaches a square Kondo lattice, and the ground state is a  $d_{x^2-y^2}$-HFSC phase in the 
metallic case. Note that in both cases, the valence bond order $\gamma$ along the diagonal bonds becomes more important and condensation amplitudes rise
when the band-filling gets lower (insets). Besides, the valence bond orders along the axial directions 
$\gamma'$ flip their sign, which is consistent with previous results.\cite{coqblin2003band}

\section{DISCUSSION AND CONCLUSION}

In the present work, we have established the emergence of superconducting instability
of conduction electrons induced by dimer ordering in the Shastry-Sutherland Kondo
lattice model and determined the conditions necessary for observing it. This is 
distinct from the one that connects the 
two superconducting states in the strong coupling limits as proposed in 
Ref.~\onlinecite{coleman2010frustration}. In order to observe the dimer-induced
heavy fermion superconducting phase, the system needs to
be doped with holes away from half-filling, whereas large values for the Kondo 
coupling $J_K$ and the hopping amplitudes are needed for the suppression  of
ordering of the localized magnetic moments as well as the formation of dimers.  

In order to destroy the PKS phase more easily, the system in the $d_{xy}$-HFSC phase must have large Kondo hybridization energy as well as hopping energy, and the $d_{xy}$-wave characteristic of the superconductivity is smeared in the 
dispersion of the low-energy excitation, as in Fig.~\hyperref[GraphK]{\ref{GraphK}(c)}. 
Nevertheless, an excitation gap can be measured at the center of the Brillouin zone. From an experimental point of view,  these effects may render the superconducting phase difficult to be realized in such materials as Yb$_2$Pt$_2$Pb whose Kondo coupling might not be strong enough.  

On the other hand, it was the large residual effect of the gauge invariance that readily led 
us to study the possibility of the PKS phase where some valence bonds remain intact. If the dimer ordering and the Kondo hybridization occur at different sites, the instability in conduction electron can not be observed.
Due to the restriction of the mean-field formalism, the PKS phase is only studied at 
$n_c =0.5$. What happens when $0.5<n_c <1$? One may expect that the PKS phase still 
survives, but it is also very likely that higher band-fillings can push the PKS-($d_{xy}$-)HFSC transition point to a smaller 
$J_K$. Whether such a phase will terminate at the transition point at half-filling is an intriguing question. Additional studies going beyond the mean-field theory are needed 
to address these issues. 

Since the dimers in the VBS phase are stable, the large-$N$ mean-field theory has a 
reliable starting point. Will the superconducting phase be disrupted by gauge field 
fluctuations? A higher
order large-$N$ expansion and a renormalization group calculation may shed light on this issue.  A recent measurement on
Yb$_2$Pt$_2$Pb indicates that the local moments on Yb atoms may have a strong Ising 
character. \cite{miiller2014magnetic} The effect of breaking SU(2) symmetry of spin correlation
on the global phase diagram is also another interesting problem. In order to answer 
these questions, more experimental and theoretical studies of the SSKL system are needed.

To summarize, we have used a more complete mean-field theory to study the zero-temperature phase diagram of the SSKL model quantitatively by explicitly including the spinon pairing of local moments. We have paid special attention to the transition between the VBS phase and the HFL phase as a combined result of both frustration and the Kondo effect. The dimer-induced $d_{xy}$-HFSC phase is shown to exist at the mean-field level when the Kondo coupling and the diagonal hopping are strong in the metallic case. A second order phase transition from the $d_{xy}$-HFSC phase to the $d_{x^2-y^2}$-HFSC phase phase is observed, with a range of coexistence. We also verify the existence of the PKS phase that precedes the superconducting phase when the dimers of the SSL lattice start to be resolved by the Kondo hybridization from conduction electrons. We hope our analysis can provide new insights into heavy fermion materials and stimulate further experimental and theoretical studies.

\begin{acknowledgments}
This work is based on L.S.'s undergraduate research at Nanyang Technological University 
with P.S..  It is a pleasure to thank Piers Coleman, Andriy Nevidomskyy and Qimiao Si for 
useful discussions and Vitor M. Pereira for valuable support during the preparation
of the manuscript. L.S. was supported by the NUS Graphene Research Centre through the Singapore NRF-CRP grant R-144-000-295-281. P.S. was supported in part by the National Science Foundation (USA)
under Grant No. PHYS-1066293 and the hospitality of the Aspen Center for Physics, as well as grant MOE2011-T2-1-108 from the Ministry of Education, Singapore. 
\end{acknowledgments}

\appendix*
\section{MEAN-FIELD HAMILTONIAN}

We give the explicit expression of the mean-field Hamiltonian here. Once we arrive at Eq.~(\ref{Eq: Hamiltonian2}), by assumption of translational symmetry and transformation to the momentum space, we can rewrite the Hamiltonian in terms of the Nambu spinors: 
\begin{equation}
{\Psi}^{\dagger}(\textbf{k}) = 
\begin{pmatrix}
 {\psi}_{c\uparrow}^{\dagger}(\textbf{k})& {\psi}_{c\downarrow} (-\textbf{k})& {\psi}_{f\uparrow}^{\dagger}(\textbf{k})& {\psi}_{f\downarrow} (-\textbf{k})
\end{pmatrix}
\end{equation}
and their conjugates ${\Psi}(\textbf{k})$
with
\begin{equation}
{\psi}_{c\uparrow}^{\dagger}(\textbf{k}) = 
\begin{pmatrix}
c_{0\uparrow}^{\dagger}({\textbf{k}}) & c_{1\uparrow}^{\dagger}({\textbf{k}})&c_{2\uparrow}^{\dagger}({\textbf{k}})&c_{3\uparrow}^{\dagger}({\textbf{k}})
\end{pmatrix} ,
\end{equation}
and
\begin{equation}
{\psi}_{f\uparrow}^{\dagger}(\textbf{k}) = 
\begin{pmatrix}
f_{0\uparrow}^{\dagger}({\textbf{k}}) & f_{1\uparrow}
^{\dagger}({\textbf{k}})&f_{2\uparrow}^{\dagger}({\textbf{k}})&f_{3\uparrow}^{\dagger}({\textbf{k}}) 
\end{pmatrix} .
\end{equation}
Algebraic manipulations give us
\begin{equation}
{\tilde{H}}_{MF} = \sum_{ \textbf{k}} {\Psi}^{\dagger}(\textbf{k}){\tilde{T}}(\textbf{k}) {\Psi} (\textbf{k})  + \tilde{E}_0 
\end{equation} 
with 
\begin{equation}
{\tilde{T}} (\textbf{k}) =
-\begin{pmatrix}
 {M} (\textbf{k})& {\bm 0} &  K(\textbf{k})& {\bm 0}   \\
 {\bm 0} & -{M} (\textbf{k}) &{\bm 0} & -K(\textbf{k})\\
 K(\textbf{k})& {\bm 0} & {N} (\textbf{k})& {\tilde{N}}^{\dagger}(\textbf{k})   \\
{\bm 0} &-K(\textbf{k})&  {\tilde{N}}(\textbf{k})  &-{N} (\textbf{k})
 \\
\end{pmatrix}
\end{equation} 
where ${M} (\textbf{k}), {N} (\textbf{k}),  {\tilde{N}}(\textbf{k})$, and $K(\textbf{k})$ are, respectively, given by
\begin{align}
M(\textbf{k})& =& \nonumber \\
&\begin{pmatrix}
 \mu  &     2 t' {\text {cos}}(\frac{x}{2})  &      2 t' {\text {cos}}(\frac{y}{2}) & t e^{i(\frac{x+y}{2})}  \\
2 t' {\text {cos}}(\frac{x}{2})  &     \mu &          t  e^{i(\frac{x-y}{2})} &  2 t' {\text {cos}}(\frac{y}{2})      \\
2 t' {\text {cos}}(\frac{y}{2}) & t e^{-i(\frac{x-y}{2})}    &         \mu &  2 t' {\text {cos}}(\frac{x}{2}) \\
  t e^{-i(\frac{x+y}{2})}   &2 t' {\text {cos}}(\frac{y}{2})  &     2 t' {\text {cos}}(\frac{x}{2})  &     \mu   \\
\end{pmatrix},  
\end{align}   
\vspace*{-10pt}
\begin{align}
N (\textbf{k}) &=& \nonumber \\
&\begin{pmatrix}
 -E_f  &     2 \Gamma '{\text {cos}}(\frac{x}{2})  &     2 \Gamma' {\text {cos}}(\frac{y}{2}) &  \Gamma e^{i(\frac{x+y}{2})} \\
2 \Gamma' {\text {cos}}(\frac{x}{2})  &     -E_f  &          \Gamma  e^{i(\frac{x-y}{2})} &  2 \Gamma' {\text {cos}}(\frac{y}{2})  \\
2 \Gamma' {\text {cos}}(\frac{y}{2})  & \Gamma e^{-i(\frac{x-y}{2})} &         -E_f  &  2 \Gamma' {\text {cos}}(\frac{x}{2}) \\
 \Gamma e^{-i(\frac{x+y}{2})}  &  2 \Gamma' {\text {cos}}(\frac{y}{2}) &     2 \Gamma' {\text {cos}}(\frac{x}{2})  &      -E_f  \\
\end{pmatrix}, 
\end{align} 
\vspace*{-10pt}
\begin{align}
&{\tilde{N}} (\textbf{k}) =& \nonumber \\
&\begin{pmatrix}
 0  &     2 B'{\text {cos}}(\frac{x}{2}) &     - 2 B'  {\text {cos}}(\frac{y}{2})  &  B  e^{i(\frac{x+y}{2})}   \\
2 B'  {\text {cos}}(\frac{x}{2}) &    0  &      -   B  e^{i(\frac{x-y}{2})} &  -2 B' {\text {cos}}(\frac{y}{2})      \\
-2 B' {\text {cos}}(\frac{y}{2})  & - B e^{-i(\frac{x-y}{2})}   &         0 &   2 B'{\text {cos}}(\frac{x}{2}) \\
 B e^{-i(\frac{x+y}{2})}   &-2 B' {\text {cos}}(\frac{y}{2})   &     2 B'  {\text {cos}}(\frac{x}{2})  &     0
\end{pmatrix},
\end{align}   
and $K(\textbf{k}) = K \textbf{I}_{4\times4}$. We have used notations $x = \textbf{k}_xa$, $y = \textbf{k}_ya$, $\Gamma = J \gamma/2$, $\Gamma' = J' \gamma'/2$, $B  = J b/2$, $B'  = J' b' /2$, $K = J_K V/2$, and  ${\tilde{E}}_0$ is given by Eq.~(\ref{Eq:tE0}). The eigenvalues of ${\tilde{T}} (\textbf{k})$ can be obtained via an extended Bogoliubov transformation.
 
\bibliographystyle{apsrev}
\bibliography{manuscript_v4}

\end{document}